\documentclass[epj]{svjour}
\usepackage{epsfig}
\usepackage{graphicx}
\usepackage{amssymb}
\begin{document}
\title{Hadronic freeze-out in an effective relativistic mean field model}
\author{A. Lavagno}
\institute{Department of Applied Science and Technology, Politecnico di Torino, I-10129 Torino, Italy and \\
Istituto Nazionale di Fisica Nucleare (INFN), Sezione di Torino, I-10126 Torino, Italy}
\date{Received: date / Revised version: date}
\abstract{We investigate an effective relativistic equation of state at finite values of temperature and baryon chemical potential with the inclusion of the full octet of baryons, the $\Delta$-isobars and the lightest pseudoscalar and vector meson degrees of freedom. These last particles have been introduced within a phenomenological approach by taking into account of an effective chemical potential and mass depending on the self-consistent interaction between baryons. In this framework, we study of the hadron yield ratios measured in central heavy ion collisions over a broad energy range and present the beam energy dependence of underlying dynamic quantities like the net baryon density and the energy density.
} 
\maketitle

\section{Introduction}

High energy heavy ion collisions provide a unique opportunity to explore in laboratory the behavior of hot and dense nuclear matter. Comparing the experimental data with different theoretical models, such as, for example, fluid-dynamical models, it is possible to study fundamental aspects of nuclear and hadron physics at different regimes of temperature and density \cite{hwa,biro2012}. This demands a detailed knowledge of the bulk thermodynamics and the equation of state (EOS) of strongly-interacting matter.

The statistical thermal model has turned out to be very successful in describing particle abundances produced in relativistic heavy ion collisions \cite{braun01,andronic06,andronic09,andronic11,becca1,becca06,cleymans06,cleymans10,cleymans12,lu02,gore}. An analysis of the energy dependence of the thermal parameters, temperature and baryon chemical potential, extracted from the fits with the experimental data, establishes essential properties of the hot and dense fireball produced at the freeze-out evolution stage, when inelastic collisions cease. Both temperature and baryon chemical potential exhibit a monotonic variation with the beam energy. As the collision energy is increased, there is an increase of
the chemical freeze-out temperature and a corresponding decrease of the baryon chemical potential.
The number of baryons becomes close to anti-baryons and meson degrees of freedom become increasingly important with a clear passage from a baryon-dominant regime to a meson-dominant one \cite{cleymans05}. In this context, it is relevant to observe that, since the noninteracting gas model neglects any kind of possible in-medium modifications, to take phenomenologically into account of the interaction between hadrons at finite densities, finite size corrections have been considered in the excluded volume approximation, in the same spirit as in a Van der Waals gas \cite{rischke,yen,mishu}.

From a more microscopic point of view, Walecka-type relativistic mean-field (RMF) models have been widely successfully used for describing the properties of finite nuclei as well as dense and finite temperature nuclear matter \cite{serot-wal,boguta,glen_plb1982,providencia2011,providencia2012,prl2007,prc2012,jpg2012}. It is
relevant to point out that such RMF models usually do not respect chiral symmetry. Furthermore, the repulsive vector field is proportional to the net baryon density, therefore, standard RMF models do not appear, in principle, fully appropriate for the very low density and high temperature regime. In this context, let us observe that a phenomenological RMF model has been recently proposed in order to calculate the EOS of hadronic matter in a broad density-temperature region by considering masses and
coupling constants depending on the $\sigma$-meson field \cite{toneev}. In that approach, motivated by the Brown-Rho scaling hypothesis, a non chiral symmetric model simulates a chiral symmetry restoration with a temperature increase. On the other hand, relativistic chiral SU(3) models have also been developed to take into account particle ratios at RHIC and baryon resonances impact on the chiral phase transition \cite{zschie,zschie2}. Such models include meson-meson and meson-baryon interactions with a consequent effective masses for all hadrons, giving a higher kinetic energy which mimics an additional repulsion among the particles. In addition, it is relevant to observe that a fit of the particle production yield ratios measured at RHIC has also been studied in the framework of a RMF model with the introduction of a phenomenological meson (repulsive) interaction corresponding to a fixed ratio $m_i^*/m_i=$1.3 (between the effective in-medium meson mass $m_i^*$ and the vacuum one $m_i$) for all considered meson particles \cite{chiapparini}.

The physical relevance of such RMF phenomenological investigations, lies in the fact that these models are able to shed light on the nature of baryon-meson and meson-meson in-medium interaction at different conditions of temperature and density and, hopefully, yield information about possible phase transition phenomena.

In Ref. \cite{prc2010}, an effective hadronic EOS has been studied by requiring the global conservation of baryon number, electric charge fraction and zero net strangeness in the range of finite temperature and density. The study has been carried out by means of an effective RMF model with the inclusion of the octet of the lightest baryons, the $\Delta$-isobar degrees of freedom and the lightest pseudoscalar and vector mesons. These last particles have been considered in the so-called one-body contribution, taking into account their effective chemical potentials depending on the self-consistent interaction between baryons. In this context, we have studied the influence of the $\Delta$-isobar degrees of freedom and, in connection, the behavior of different particle-antiparticle ratios and strangeness production.

In this paper we are going to include in a phenomenological way the contribution of the mean field interaction in the effective (in-medium) meson masses in order to study the hadron production yield ratios and compare our results with the experimental central nucleus-nucleus collisions, from AGS to RHIC, in the energy range $\sqrt{s_{NN}}=2.7 \div 200$ GeV.


\section{The effective relativistic mean field model}

The total Lagrangian density ${\mathcal L}$ can be written as
\begin{equation}
{\mathcal L}={\mathcal L}_{\rm octet}+{\mathcal L}_\Delta+{\mathcal
L}_{\rm qpm} \, ,
\end{equation}
where ${\mathcal L}_{\rm octet}$ stands for the full octet of the
lightest baryons ($p$, $n$, $\Lambda$, $\Sigma^+$, $\Sigma^0$,
$\Sigma^-$, $\Xi^0$, $\Xi^-$) interacting with $\sigma$, $\omega$,
$\rho$ meson fields; ${\mathcal L}_\Delta$ corresponds
to the degrees of freedom for the $\Delta$-isobars ($\Delta^{++}$,
$\Delta^{+}$, $\Delta^0$, $\Delta^-$) and ${\mathcal L}_{\rm qpm}$
is related to a quasi-particle gas of the lightest pseudoscalar and
vector mesons with effective chemical potentials and masses (see below for details).

The Lagrangian for the self-interacting octet of baryons can
be written as \cite{serot-wal,boguta,glen_plb1982}
%
\begin{eqnarray}\label{lagrangian}
{\mathcal L}_{\rm octet} &=&
\sum_k\overline{\psi}_k\,[i\,\gamma_{\mu}\,\partial^{\mu}-(M_k-
g_{\sigma k}\,\sigma) -g_{\omega
k}\,\gamma_\mu\,\omega^{\mu} \nonumber\\
&&-g_{\rho k}\,\gamma_{\mu}\,\vec{t}
\cdot \vec{\rho}^{\;\mu}]\,\psi_k+\frac{1}{2}(\partial_{\mu}\sigma\partial^{\mu}\sigma-m_{\sigma}^2\sigma^2)
\nonumber\\
&& -\frac{1}{3}a\,(g_{\sigma
N}\,\sigma)^{3}-\frac{1}{4}\,b\,(g_{\sigma N}\,\sigma^{4})+\frac{1}{2}\,m^2_{\omega}\,\omega_{\mu}\omega^{\mu}
\nonumber\\
&&+\frac{1}{4}\,c\,(g_{\omega N}^2\,\omega_\mu\omega^\mu)^2+\frac{1}{2}\,m^2_{\rho}\,\vec{\rho}_{\mu}\cdot\vec{\rho}^{\;\mu}\nonumber\\
&&-\frac{1}{4}F_{\mu\nu}F^{\mu\nu}-\frac{1}{4}\vec{G}_{\mu\nu}\vec{G}^{\mu\nu}\,,
\end{eqnarray}
where the sum runs over the full octet of baryons, $M_k$ is the
vacuum baryon mass of index $k$, the quantity $\vec{t}$ denotes
the isospin operator that acts on the baryon.

In the regime of finite values of temperature and density, a state of high density resonance matter may be formed and the $\Delta(1232)$-isobar degrees of freedom are expected to play a central role \cite{zabrodin}. In particular, the formation of resonances contributes essentially to enhanced strangeness, baryon stopping and hadronic flow effects \cite{mattiello}. Transport model calculations and experimental results indicate that an excited state of baryonic matter is dominated by the $\Delta$-resonance at the energy from AGS to RHIC \cite{hofmann,bass,mao,schaffner91,fachini,star-res}.
To incorporate $\Delta$-isobars in the framework of effective hadron field theories, a formalism was developed to treat
$\Delta$ analogously to the nucleon, taking only the on-shell $\Delta$s into account and the mass of the $\Delta$s are substituted by the effective one in the RMF approximation \cite{malfliet,boguta82}.

The Lagrangian density concerning the $\Delta$-isobars can then be expressed as \cite{boguta82,greiner97,kosov}
\begin{eqnarray}
{\mathcal L}_\Delta=\overline{\psi}_{\Delta\,\nu}\, [i\gamma_\mu
\partial^\mu -(M_\Delta-g_{\sigma\Delta}
\sigma)-g_{\omega\Delta}\gamma_\mu\omega^\mu
 ]\psi_{\Delta}^\nu \,
,
\end{eqnarray}
where $\psi_\Delta^\nu$ is the Rarita-Schwinger spinor for the $\Delta$-baryon. Due to the uncertainty on the
meson-$\Delta$ coupling constants, we limit ourselves to consider
only the coupling with $\sigma$ ($g_{\sigma\Delta}$) and $\omega$ ($g_{\omega\Delta}$) meson fields, more of which are explored in the literature \cite{greiner97,kosov,jin}.

In the RMF approach, baryons are considered as Dirac
quasiparticles moving in classical meson fields and the field
operators are replaced by their expectation values. In this
context, it is relevant to remember that the RMF model does not
respect chiral symmetry and the contribution coming from the
Dirac-sea and the quantum fluctuation of the meson fields are
neglected. As a consequence, the field equations in RMF
approximation have the following form
\begin{eqnarray}
&&\Big(i\gamma_\mu\partial^\mu-M_k^*-g_{\omega
k}\gamma^0\omega-g_{\rho k}\gamma^0 t_{3 k}\rho\Big)\psi_k=0 \, ,\label{bfield}\\
&&\Big(i\gamma_\mu\partial^\mu-M_\Delta^*-g_{\omega
\Delta}\gamma^0\omega\Big)\psi_\Delta^\nu=0 \, ,\label{dfield}\\
&&m_{\sigma}^2\sigma+ a\,g_{\sigma N}^3\,{\sigma}^2+
b\,g_{\sigma N}^4\,{\sigma}^3=\sum_i g_{\sigma i}\, \rho_i^S\, , \label{sfield}\\
&&m^2_{\omega}\omega+c\,g_{\omega N}^4\,\omega^3=\sum_i g_{\omega i}\,\rho_i^B \, ,\label{ofield} \\
&&m^2_{\rho}\rho=\sum_i g_{\rho i} \,t_{3 i}\,\rho_i^B\, ,
\label{rfield}
\end{eqnarray}
where $\sigma=\langle\sigma\rangle$,
$\omega=\langle\omega^0\rangle$ and $\rho=\langle\rho^0_3\rangle$
are the nonvanishing expectation values of meson fields. The
effective mass of $i$-th baryon is given by
\begin{equation}
M_i^*=M_i-g_{\sigma i}\sigma\, .\label{mneff}
\end{equation}
In the meson-field equations,
Eqs.(\ref{sfield})-(\ref{rfield}), the sums run over all
considered baryons (octet and $\Delta$s) and $\rho_i^B$ and
$\rho_i^S$ are the baryon density and the scalar density of the
particle of index $i$, respectively. They are given by
\begin{eqnarray}
&&\rho_i^B=\gamma_i \int\frac{{\rm d}^3k}{(2\pi)^3}\;[f_i(k)-\overline{f}_i(k)] \, , \\
&&\rho_i^S=\gamma_i \int\frac{{\rm
d}^3k}{(2\pi)^3}\;\frac{M_i^*}{E_i^*}\; [f_i(k)+\overline{f}_i(k)]
\, ,
\end{eqnarray}
where $\gamma_i=2J_i+1$ is the degeneracy spin factor of the
$i$-th baryon ($\gamma_{\rm octet}=2$ for the baryon octet and
$\gamma_\Delta=4$) and $f_i(k)$ and $\overline{f}_i(k)$ are the
fermion particle and antiparticle distributions
\begin{eqnarray}
&&           f_i(k)=\frac{1}{\exp\{(E_i^*(k)-\mu_i^*)/T\}+1} \, , \\
&&\overline{f}_i(k)=\frac{1}{\exp\{(E_i^*(k)+\mu_i^*)/T\}+1} \, .
\end{eqnarray}
The effective chemical
potentials $\mu_i^*$ are given in terms of the chemical potentials
$\mu_i$ by means of the following relation
\begin{equation}
\mu_i^*=\mu_i-g_{\omega i}\,\omega-g_{\rho i}\,t_{3 i}\,\rho\, ,
\label{mueff}
\end{equation}
where $t_{3 i}$ is the third component of the isospin of $i$-th
baryon. The baryon effective energy is defined as
$E_i^*(k)=\sqrt{k^2+{{M_i}^*}^2}$.

Because we are going to describe the nuclear EOS at finite temperature
and density with respect to strong interaction, we have to
require the conservation of three "charges": baryon number (B),
electric charge (C) and strangeness number (S). Each conserved
charge has a conjugated chemical potential and the system is
described by three independent chemical potentials: $\mu_B$,
$\mu_C$ and $\mu_S$. Therefore, the chemical potential of particle
of index $i$ can be written as
\begin{equation}
\mu_i=b_i\, \mu_B+c_i\,\mu_C+s_i\,\mu_S \, , \label{mu}
\end{equation}
where $b_i$, $c_i$ and $s_i$ are, respectively, the baryon, the
electric charge and the strangeness quantum numbers of $i$-th
hadronic species.

The thermodynamical quantities can be obtained from the grand
potential $\Omega_B$ in the standard way. More explicitly, the
baryon pressure $P_B=-\Omega_B/V$ and the energy density
$\epsilon_B$ can be written as
\begin{eqnarray}
P_B&=&\frac{1}{3}\sum_i \,\gamma_i\,\int \frac{{\rm
d}^3k}{(2\pi)^3} \;\frac{k^2}{E_{i}^*(k)}\;
[f_i(k)+\overline{f}_i(k)] -\frac{1}{2}\,m_\sigma^2\,\sigma^2 \nonumber \\
&-& \frac{1}{3}a\,(g_{\sigma
N}\,\sigma)^{3}-\frac{1}{4}\,b\,(g_{\sigma N}\,\sigma^{4})+
\frac{1}{2}\,m_\omega^2\,\omega^2\nonumber\\
&+&\frac{1}{4}\,c\,(g_{\omega
N}\,\omega)^4+\frac{1}{2}\,m_{\rho}^2\,\rho^2 \, ,\\
\epsilon_B&=&\sum_i \,\gamma_i\,\int \frac{{\rm
d}^3k}{(2\pi)^3}\;E_{i}^*(k)\; [f_i(k)+\overline{f}_i(k)]
+\frac{1}{2}\,m_\sigma^2\,\sigma^2\nonumber \\
&+&\frac{1}{3}a\,(g_{\sigma
N}\,\sigma)^{3}+\frac{1}{4}\,b\,(g_{\sigma
N}\,\sigma^{4})+\frac{1}{2}\,m_\omega^2\,\omega^2\nonumber\\
&+&\frac{3}{4}\,c\,(g_{\omega
N}\,\omega)^4 +\frac{1}{2}\,m_{\rho}^2 \,\rho^2 \,  .
\end{eqnarray}

\subsection{The baryon couplings}

The numerical evaluation of the above quantities can be performed
if the meson-nucleon, -$\Delta$ and -hyperon coupling
constants are known. Concerning the nucleon coupling
constants ($g_{\sigma N}$, $g_{\omega N}$, $g_{\rho N}$),
they are determined to reproduce the bulk properties of
equilibrium nuclear matter such as the saturation densities, the
binding energy, the symmetric energy coefficient, the compression
modulus and the effective Dirac mass at saturation. Because of a
valuable range of uncertainty in the empirical values that must be
fitted, especially for the compression modulus and for the
effective Dirac mass, in literature there are different sets of
coupling constants. In this investigation we consider two
different parameter sets: the set marked TM1, from Ref.\cite{toki}, and GM3, from Glendenning and Moszkowski \cite{glen_prl1991}. The above EOSs are compatible with intermediate heavy ion collisions
constraints and extensively used in various high density astrophysical applications \cite{prc2010,daniele,ditoro,epja2009,epja2011,astro1,astro2,astro3}.

In this context let us observe that in Ref.\cite{prc2010} also the NL$\rho\delta$ EOS was considered.
This last parameter set contains in addition the $\delta$-isovector-scalar meson-baryon couplings
and have not been considered here because the GM3 and NL$\rho\delta$ EOS have comparable behaviors,
due to similar values of the compression modulus and the effective nucleon mass \cite{prc2010,toki,glen_prl1991}.
Moreover, $\delta$-meson should have a minor role in the analysis of the hadronic freeze-out
yield ratios, with lower baryon density involved and approximately constant
electric charge fraction.

As regards the implementation of hyperon degrees of freedom, they
come from determination of the corresponding meson-hyperon coupling constants that have been fitted to hypernuclear
properties \cite{schaffner_prl1993}. The adopted coupling constants and the vacuum masses are the same as in Ref. \cite{prc2010}.


%
%

Concerning the formation of $\Delta$-isobar matter at finite
temperature and density, it has been predicted that a phase transition from
nucleonic matter to $\Delta$-excited nuclear matter can take place
and the occurrence of this transition sensibly depends on the
$\Delta$-meson fields coupling constants \cite{greiner97}.
Referring to QCD finite-density sum rule results, which predict
that there is a larger net attraction for a $\Delta$-isobar than
for a nucleon in the nuclear medium \cite{jin}, the range of
values for the $\Delta$ coupling constants has been confined
within a triangle relation \cite{kosov}. Therefore, in setting the following coupling ratios:
\begin{equation}
 x_{\sigma\Delta}=\frac{g_{\sigma\Delta}}{g_{\sigma N}} \ \ \ {\rm and} \ \ \ x_{\omega\Delta}=\frac{g_{\omega\Delta}}{g_{\omega N}} \, ,
\end{equation}
we have to require
that i) the second minimum of the energy per
baryon lies above the saturation energy of normal nuclear matter,
i.e., in the mixed $\Delta$-nucleon phase only a metastable state
can occur; ii) there are no $\Delta$-isobars present at the
saturation density; iii) the scalar field is more (equal)
attractive and the vector potential is less (equal) repulsive for
$\Delta$s than for nucleons, in accordance with QCD finite-density
calculations \cite{jin}. Of course, the choice of couplings that
satisfies the above conditions is not unique but there exists a finite
range of possible values (represented as a triangle region in the
plane $x_{\sigma\Delta}$--$x_{\omega\Delta}$) which depends on the
particular EOS under consideration \cite{prc2010,kosov}. Without loss of
generality, we can limit our investigation to move only in a side
of such a triangle region by fixing $x_{\omega\Delta}=1$ and varying $x_{\sigma\Delta}$ from unity to a maximum value compatible with the conditions mentioned
above. Comparable conclusions are obtained with any other compatible
choice of the two coupling ratios (see Ref.\cite{prc2010} for a more detailed discussion about the value of the $\Delta$-couplings and the formation of $\Delta$ metastable matter for different EOSs).
In this investigation, we fix the scalar coupling ratio to the value
$x_{\sigma\Delta}=$1.25 in the TM1 and $x_{\sigma\Delta}=$1.40 in the GM3 parameter set, for which the $\Delta$ metastable condition is not realized. In addition, we consider the case of  $x_{\sigma\Delta}=$1.33 in the TM1 and $x_{\sigma\Delta}=$1.50 in the GM3 parameter set, corresponding to the maximum value in which $\Delta$ metastable state can be realized \cite{prc2010}. Therefore, we consider two very different scenarios (not metastable and metastable $\Delta$-matter) for the two different parameter sets (note that, when a metastable state is not realized, decays rate
are not taken into account in this approach).

\subsection{Meson degrees of freedom and chemical equilibrium}
It is well known that the lightest pseudoscalar and vector mesons play a crucial role in the EOS, especially at low baryon
density and high temperature.
On the other hand, the contribution of the $\pi$ mesons (and other
pseudoscalar and pseudovector fields) vanishes at the mean-field
level. From a phenomenological point of view, we can take into
account the meson particle degrees of freedom by adding their
one-body contribution to the thermodynamical potential, that is,
the contribution of a quasi-particle Bose gas with an effective chemical
potential $\mu_j^*$ and an effective mass $m_j^*$ for the $j$-meson, which
contain the self-consistent interaction of the meson fields (see next subsections for details).
Following this working hypothesis, we can evaluate the pressure
$P_M$, the energy density $\epsilon_M$ and the particle density
$\rho_j^M$ of mesons as
\begin{eqnarray}
&&P_M= \frac{1}{3}\sum_j \,\gamma_j\,\int \frac{{\rm
d}^3k}{(2\pi)^3} \;\frac{k^2}{E_{j}(k)}\; g_j(k)\, ,
\label{pmeson}\\
&&\epsilon_M=\sum_j \,\gamma_j\,\int \frac{{\rm
d}^3k}{(2\pi)^3}\;E_{j}(k)\; g_j(k) \, ,\label{emeson}\\
&&\rho_j^M=\gamma_j \int\frac{{\rm d}^3k}{(2\pi)^3}\;g_j(k) \, ,
\label{rhomeson}
\end{eqnarray}
where $\gamma_j=2J_j+1$ is the degeneracy spin factor of the
$j$-th meson ($\gamma=1$ for pseudoscalar mesons and $\gamma=3$
for vector mesons), the sum runs over the lightest pseudoscalar
mesons ($\pi$, $K$, $\eta$, $\eta'$) and the
lightest vector mesons ($\rho$, $\omega$, $K^*$, $\phi$),
considering the contribution of particle and antiparticle separately. In Eqs.(\ref{pmeson})-(\ref{rhomeson}) the function
$g_j(k)$ is the boson particle distribution given by
\begin{equation}
g_j(k)=\frac{1}{\exp\{(E_j^*(k)-\mu_j^*)/T\}-1} \, ,
\end{equation}
where $E_j^*(k)=\sqrt{k^2+{{m_j}^{*}}^2}$. The corresponding
antiparticle distribution will be obtained
with the substitution $\mu_j^* \rightarrow -\mu_j^*$.
Moreover, the boson integrals are subjected to the constraint $\vert\mu_j^*\vert\le m_j^*$, otherwise Bose condensation
becomes possible. We have verified that such a condition is never realized in the considered range of temperature and density.

All the aforementioned equations must
be evaluated self-consistently fulfilling the chemical equilibrium and by imposing the constraints of baryon number, electric charge and strangeness. Therefore, at a given temperature $T$, baryon density $\rho_B$, net electric charge fraction $Z/A$ ($\rho_C=Z/A\,\rho_B$) and zero net
strangeness of the system ($\rho_S=0$), the chemical potentials $\mu_B$, $\mu_C$ and $\mu_S$ are univocally determined by the following equations
\begin{eqnarray}
&&\rho_B=\sum_i b_i\,\rho_i(T,\mu_B,\mu_C,\mu_S) \, ,\\
&&\rho_C=\sum_i c_i\, \rho_i(T,\mu_B,\mu_C,\mu_S) \, , \\
&&\rho_S=\sum_i s_i\,\rho_i(T,\mu_B,\mu_ C,\mu_S) \, ,
\end{eqnarray}
where the sums run over all considered particles (baryons and
mesons).

Finally, the total pressure and energy density will be
\begin{eqnarray}
&&P=P_B+P_M \, ,\\
&&\epsilon=\epsilon_B+\epsilon_M \, .
\end{eqnarray}

\subsubsection{Effective meson chemical potentials}
Following Refs. \cite{prc2010,muller,jpg2010}, the values of the effective
meson chemical potentials $\mu_j^*$ are obtained from the "bare"
ones $\mu_j$, given in Eq.(\ref{mu}), and subsequently expressed
in terms of the corresponding effective baryon chemical
potentials, respecting the strong interaction. For example, we
have from Eq.(\ref{mu}) that
$\mu_{\pi^+}=\mu_{\rho^+}=\mu_C\equiv\mu_p-\mu_n$ and the
corresponding effective chemical potential can be written as
\begin{eqnarray}
\mu_{\pi^+(\rho^+)}^*&\equiv&\mu_p^*-\mu_n^*
=\mu_{\pi^+(\rho^+)}-g_{\rho N}\,\rho \, , \label{mueff_m1}
\end{eqnarray}
where the last equivalence follows from Eq.(\ref{mueff}).

Analogously, by setting $x_{\omega \Lambda}=g_{\omega
\Lambda}/g_{\omega N}$, we have
\begin{eqnarray}
\!\!\!\!\!\!\!\!\!\mu_{K^+(K^{*+})}^* &\equiv& \mu_p^*-\mu_{\Lambda(\Sigma^0)}^*\nonumber\\
&=&\mu_{K^+(K^{*+})}- (1-x_{\omega \Lambda})g_{\omega N }\omega-
\frac{1}{2}g_{\rho N}\rho \, ,\label{mueff_m2}\\
\!\!\!\!\!\!\!\!\!\mu_{K^0(K^{*0})}^* &\equiv& \mu_n^*-\mu_{\Lambda(\Sigma^0)}^*\nonumber\\
&=&\mu_{K^0(K^{*0})}- (1-x_{\omega \Lambda})g_{\omega N }\omega+
\frac{1}{2}g_{\rho N}\rho \, , \label{mueff_m3}
\end{eqnarray}
while the others strangeless neutral mesons have a vanishing
chemical potential.
\subsubsection{Effective meson masses}

Besides the chemical potentials, the mean field interaction
also affects the values of the effective meson masses. Here, we are going to
introduce an analogue phenomenological approach to that used for
the particle chemical potentials in order to include, at different values of temperature and baryon density, the in-medium interaction in the effective meson masses.

As well as the effective meson chemical potentials have been obtained from
a difference between the effective baryon chemical potentials,
so we postulate that the meson effective masses can be expressed as
a difference between the effective baryon masses respecting the strong
interaction and the main processes of meson production/absorption involving different baryons. More
explicitly, concerning pions, being the $\Delta$-isobar state one of the
most prominent feature of $\pi N$ dynamics, the main process involving different baryons can be identified with the
$\Delta\leftrightarrow\pi N$ one. Hence, from a phenomenological point of view, we can postulate the variation of the effective pion mass $m_\pi^*$, with respect to the vacuum one $m_\pi$, in terms of the variation between the $\Delta$ and nucleon effective masses, with respect to their vacuum ones. With this assumption, we can write the effective pion mass as
\begin{equation}
m_\pi^*=m_\pi-(x_{\sigma\Delta}-1) g_{\sigma N}\,\sigma \, . \label{mpieff}
\end{equation}

Because we set $x_{\sigma\Delta}>1$, the above equation
requires a reduction of the effective pion mass and such a
reduction depends on the value of the $\sigma$ meson field
related to the self-consistent interaction between hadrons. Let us
remark that Eq.(\ref{mpieff}) does not imply that the production/aborption of
pions is due to the formation of $\Delta$-particles but rather
that the self-consistent mean field interaction is phenome\-no\-logically related to the
difference $g_{\sigma\Delta}-g_{\sigma N}$ between the $\Delta$ and the nucleon $\sigma$-field
couplings.

In this context, it is proper to observe that chiral model calculations predict an enhancement of the effective pion mass at RHIC energies \cite{zschie}. On the other hand, the in-medium pion dispersion relation in a relativistic quantum transport theory implies a smaller effective pion mass in cold and dense nuclear matter \cite{mao1999}.

Let us further outline that from deeply bound pionic states in nuclei and $\pi$-atoms, for symmetric nuclear matter at zero temperature and at the saturation nuclear density, different investigations indicate that the effective pion mass results to be very close to the free one \cite{weise1997,weise,migdal1990}. From Eq.(\ref{mpieff}), this implies $x_{\sigma\Delta}\approx 1$ at $T=0$ and $\rho_B=\rho_0$. In principle, such a condition could be accomplished by introducing an in-medium dependence of the $\Delta$-meson coupling constant as considered, for example, in Ref. \cite{mao1999}. On the other hand, to study the nuclear EOS in the full range of temperature and density is far beyond the scope of this investigation. Because we are mainly interested in a phenomenological analysis at finite temperatures reachable in high energy heavy ion collisions, far from the saturation nuclear matter regime, in order to avoid further degrees of freedom, we will limit ourselves to considering a fixed value of the $x_{\sigma\Delta}$ coupling.


Following the above scheme, we are able to define in a similar manner all the other
effective meson masses. Concerning the other strangeless mesons, from the point of view of the effective mean field interaction, they can be considered as correlated states of pions in the nuclear
medium having the corresponding dependence of the effective meson
masses. For example, for the $\rho$ meson, we can assume the process: $\rho\leftrightarrow
2\,\pi$ and for the $\omega$ meson: $\omega\leftrightarrow 3\,\pi$.
Therefore, we have, respectively,
\begin{eqnarray}
&&m_\rho^*=m_\rho-\,2\,(x_{\sigma\Delta}-1) g_{\sigma N}\,\sigma
\,, \\
&&m_\omega^*=m_\omega-\,3\,(x_{\sigma\Delta}-1) g_{\sigma
N}\,\sigma \,,
\end{eqnarray}
where we have used the result given in Eq.(\ref{mpieff}). \\
On the other hand, because the $\eta$ meson is not allowed to strongly
decay into pions, for $\eta$ and $\eta'$ mesons we limit to consider
their respective vacuum masses.

Considering the strong interaction only, the effective kaon masses will be related to the difference of the effective hyperon and nucleon masses mainly by means of two different channels \cite{wandapr}\footnote{Of course there are other important channels in the kaon production/absorption process, such as, for example, $\pi\bar{\pi}\leftrightarrow K \bar{K}$. However, because this last one does not involve different baryon particles, cannot be considered in our phenomenological scheme related to the determination of the effective meson masses. On the other hand, kaon photo- or electro-production cannot be taken into account in our EOS.}: the associate production/absorption due to pion
conversion modes on a single nucleon
\begin{eqnarray}
\pi N\leftrightarrow \Lambda K \, , \ \ \ \pi N\leftrightarrow
\Sigma K \, ,\ \ \ \pi N\leftrightarrow \Xi KK \, , \label{ch1}
\end{eqnarray}
the channel due to non-pionic modes on two nucleons
\begin{eqnarray}
\!\!\!\!\!\!\!\!N N\leftrightarrow N\Lambda K\, , \ \ \ N N\leftrightarrow N\Sigma
K\, , \ \ \ N N\leftrightarrow N\Xi KK \, , \label{ch2}
\end{eqnarray}
and any conjugate process involving the same type of particle/antiparticle \footnote{It is proper to observe that in this approach we are not taking into account the asymmetry between particle and antiparticle effective meson masses, neglecting, for example, possible repulsive potential for kaons and attractive for antikaons, especially relevant in a low temperature regime \cite{gerry,mishra}.}.

In the literature there is uncertainty about which of the above two channels is dominant in a nuclear medium at different values of temperature and density \cite{mares}.
Taking into account that we are going to study the EOS in regime of warm or hot nuclear matter, where mesons become dominant on the baryon degrees of freedom, for simplicity in the following we will limit our considerations to the first channel only.

As before, the guiding phenomenological principle that we follow in the determination of the effective mass of kaons (antikaons) is related to the different $\sigma$ field interaction between the different baryons involved into the production/absoption process.
Considering the processes indicated in Eq.(\ref{ch1}), kaons are always related to the presence of
hyperons, therefore, being the scalar $\sigma$ field less attractive for hyperons than for nucleons ($x_{\sigma Y}=g_{\sigma Y}/g_{\sigma N}<1$) \cite{prc2010}, we can postulate an increase in the effective kaon mass $m_K^*$. Following the above criterion, we can set the kaon effective mass as follows
\begin{equation}
m_K^*=m_K+\left[x_{\sigma\Delta}-\frac{x_{\sigma
\Lambda}+x_{\sigma \Sigma}}{2}\right] \, g_{\sigma N}\,\sigma\, ,
\label{mkeff}
\end{equation}
where we have taken the average contribution between the first two modes of Eq.(\ref{ch1}),
neglecting the last one involving multistrange production/absorption and we have used Eq.(\ref{mpieff}) for the effective pion mass.

Moreover, concerning the effective mean field interaction, $K^*$ meson can be viewed as a strongly correlated state
of $K$ and $\pi$ ($K^*\leftrightarrow K\pi$) and its effective
mass will be expressed as
\begin{equation}
m_{K^*}^*=m_{K^*}+\left[1-\frac{x_{\sigma \Lambda}+x_{\sigma
\Sigma}}{2}\right]\, g_{\sigma N}\,\sigma\, ,
\end{equation}
where we have made explicit the effective pion and kaon masses given in eqs. (\ref{mpieff}) and (\ref{mkeff}), respectively.

Finally, according to the Zweig rule,  $\phi$ meson decays mainly into two kaons ($\phi\leftrightarrow 2\,K$), therefore, its effective mass can be written as follows
\begin{equation}
m_\phi^*=m_\phi+2\,\left[x_{\sigma\Delta}-\frac{x_{\sigma
\Lambda}+x_{\sigma \Sigma}}{2}\right] \, g_{\sigma N}\,\sigma\, ,
\label{meffphi}
\end{equation}
where we have used the effective kaon mass given in Eq.(\ref{mkeff}).

Summarizing, in this subsection we have postulated the meson effective masses as a difference between the effective masses of different baryons involved in the meson production/absorption strong processes.  The correction to the vacuum masses is driven by the $\sigma$-meson field obtained self-consistently at different values of temperature and baryon density. In this scheme, the meson effective mass $m_j^*$ of species $j$ results to be reduced (attractive interaction) or enhanced (repulsive interaction) depending if the coupling $g_{\sigma B}$ with the heavier baryon involved is greater (as in the case of $\pi$, $\rho$ and $\omega$ mesons) or lower (as in the case of $K$, $K^*$ and $\phi$ mesons) than the nucleon one ($g_{\sigma N}$). Although the above new phenomenological assumption introduced in this paper
has a certain degree of arbitrariness, it has the evident advantage of not introducing additional
parameters/couplings and, as we will see in the next section,  will play a
crucial role in the determination of the particle ratios at different energy range




\section{Hadron yield ratios for central nucleus-nucleus collisions}

In this Section, we are going to compare the hadronic ratios,
obtained in the framework of the effective EOS, with the yield ratios measured in central high energy heavy ion collisions from AGS to RHIC energy range. At this scope, as in
statistical thermal models, we consider the temperature and the
baryon chemical potential as free parameters of the hadronic EOS (we do not consider proper volume corrections and strangeness suppression factor) and we adjust them by fitting the data minimizing the $\chi^2$ distribution \cite{andronic06,becca06,cleymans06}
\begin{equation}
\chi^2=\sum_i \left(\frac{R_i^{\rm exp}-R_i^{\rm eos
}}{\sigma_i}\right)^2 \, ,
\end{equation}
where $R_i^{\rm exp}$ is the measure of the {\it i}th ratio
yields with its relative uncertainty $\sigma_i$ (sum in quadrature of statistical and systematic experimental errors) and $R_i^{\rm eos}$ is the corresponding value obtained from the effective hadronic EOS.

It is obvious that the chemical freeze-out cannot correctly be
described as a mixture of the lightest baryons and mesons
considered in the above EOS without properly considering decays,
rescattering, annihilation effects and, eventually, nonequilibrium processes.
Furthermore, in the considered EOS the particles are off-shell
(with an effective mass) and, in the determination of the hadronic ratios, the effects of any mechanism to bring them on-shell are not taken into account \cite{schaffner91}. Therefore, the obtained
temperatures and baryon chemical potentials cannot properly be considered as the freeze-out parameters. However, as in Refs.  \cite{zschie,chiapparini}, such a fitting procedure can give interesting indications on the nature of the in-medium nuclear interaction at finite temperature and density regime.

\begin{figure*}
\begin{center}
\resizebox{0.8\textwidth}{!}{%
\includegraphics{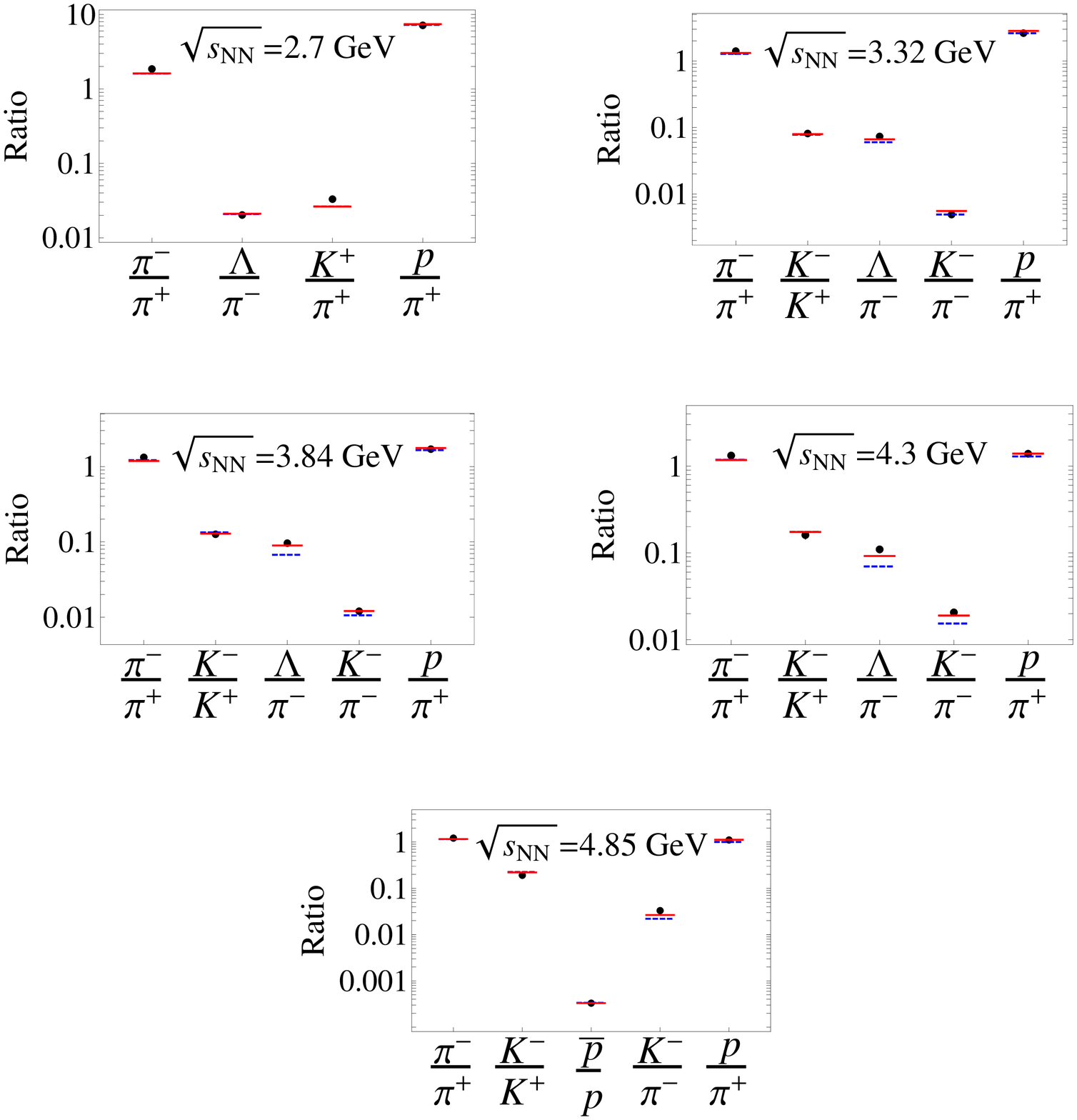}}
\caption{Measured hadron yield ratios compared to the effective
mean field model results at the AGS beam energies. The dashed
lines correspond to the TM1 parameter set with $x_{\sigma\Delta}=1.33$ and the solid lines to
the GM3 one with $x_{\sigma\Delta}=1.40$.} \label{ags}
\end{center}
\end{figure*}

Taking into account the above observations, we limit ourselves to study only hadron yield ratios (therefore the enclosed volume of the system plays no role in the analysis) and, for simplicity, to consider a limited set of experimental ratios. In order to make such a choice as little as possible arbitrary, we include in the fit, for SPS and RHIC data, the same five independent yield ratios considered
in the recent data analysis of Refs. \cite{star_prc09,qm2011}: $\pi^-/\pi^+$, $K^-/K^+$, $\overline{p}/p$, $K^-/\pi^-$, $\overline{p}/\pi^-$ and, at the AGS energies, we choose the most similar set of data (see Fig. \ref{ags}) on the basis of the experimental ratios considered in Ref. \cite{andronic06}. Therefore, except in the case of beam energy $\sqrt{s_{NN}}=130$ GeV (see below for details), we refer to \cite{andronic06} and references therein for the AGS and SPS midrapidity data and to \cite{star_prc09} for the RHIC midrapidity data. In the central Au-Au collisions at AGS and RHIC, the electric charge fraction has been fixed to $Z/A=0.401$, while in the SPS Pb-Pb collisions $Z/A=0.394$. As already stated, we require a zero net strangeness at all energies, therefore, at this stage the analysis does not consider the possible presence of a non-vanishing net strangeness at midrapidity \cite{steinheimer}.

\begin{figure*}
\begin{center}
\resizebox{0.8\textwidth}{!}{%
\includegraphics{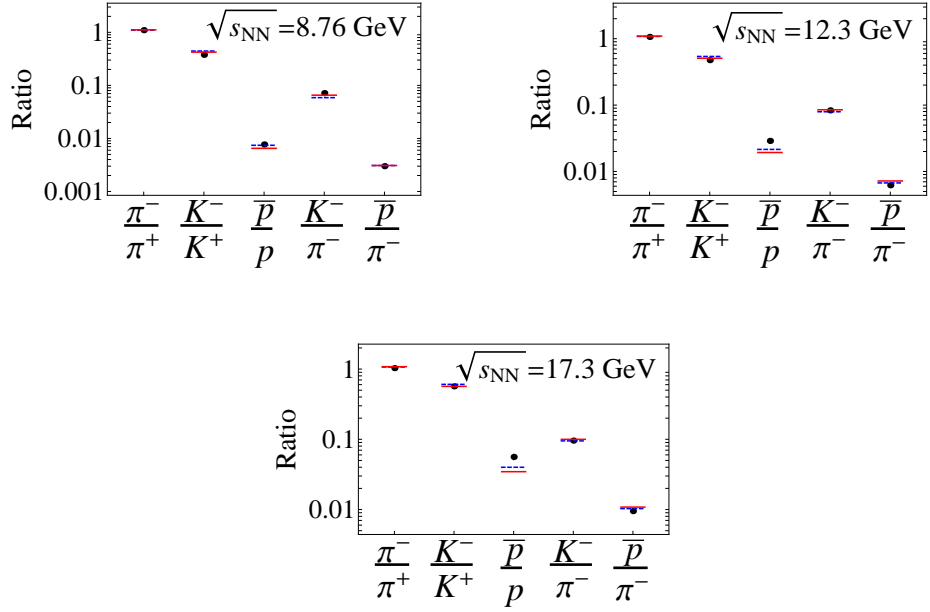}}
\caption{The same of Fig.\ref{ags} at the SPS beam energies.}
\label{sps}
\end{center}
\end{figure*}

\begin{figure*}
\begin{center}
\resizebox{0.8\textwidth}{!}{%
\includegraphics{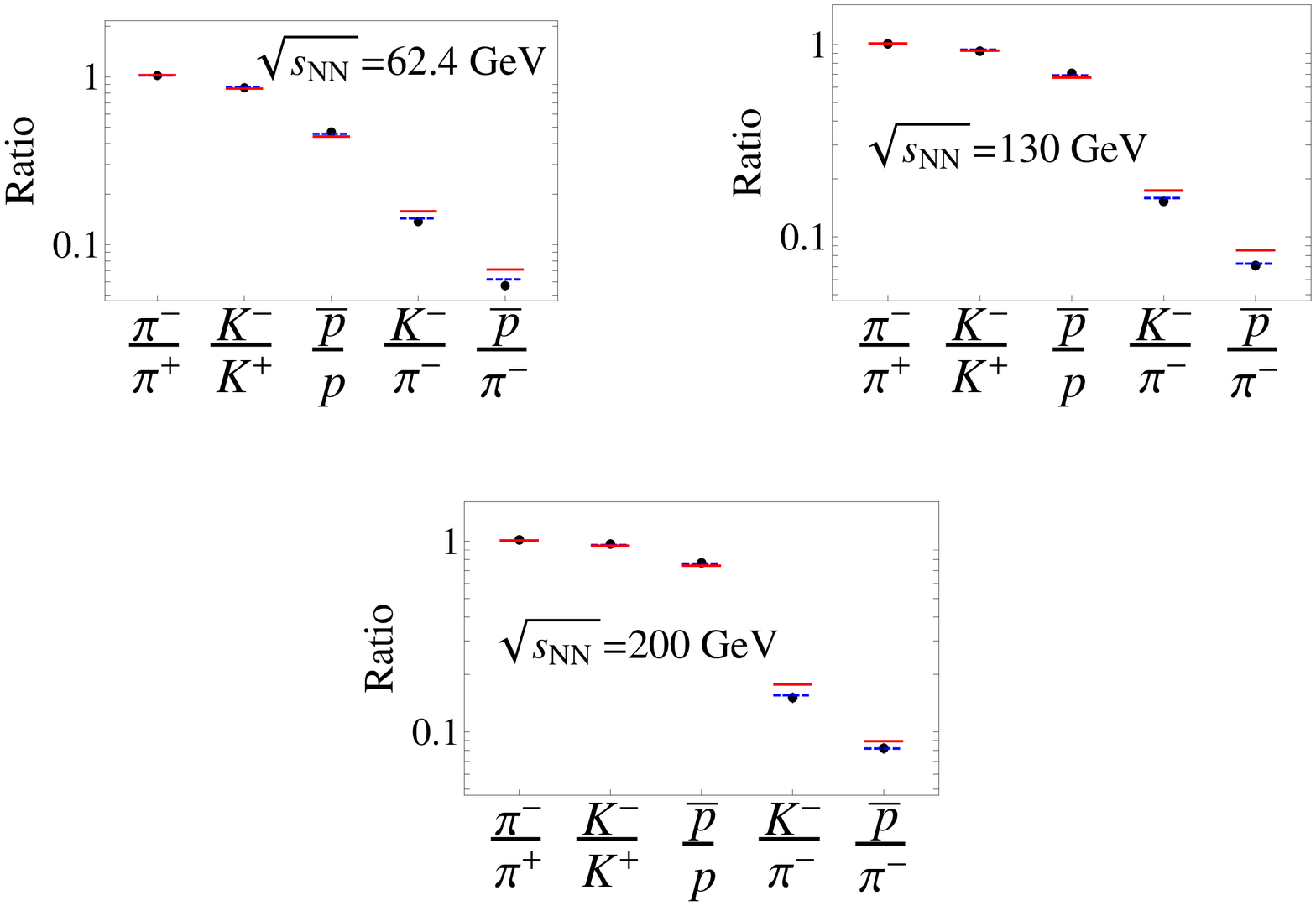}}
\caption{The same of Fig.\ref{ags} at the RHIC beam energies.}
\label{rhic}
\end{center}
\end{figure*}

The comparisons between the measured and calculated ratios for the best fit are reported in Figs. \ref{ags}, \ref{sps} and \ref{rhic} for the AGS, SPS and RHIC beam energies, respectively (for clarity, in the figures we report only the results with $x_{\sigma\Delta}=1.33$ for the TM1 parameter set and with $x_{\sigma\Delta}=1.40$ for the GM3 set).

In Table \ref{table_summary_tm1}, for the TM1 parameter set, and in Table \ref{table_summary_gm3}, for the GM3 parameter set, we summarize the complete results for the best fits reporting, for any considered beam energy, the obtained free parameters $T$ and $\mu_B$ with the corresponding $\chi^2_{\rm dof}=\chi^2/N_{\rm dof}$ (where $N_{\rm dof}=2$ at $\sqrt{s_{NN}}=2.7$ GeV and $N_{\rm dof}=3$ in the other cases). For comparison, in the round brackets are reported the results without considering effective meson chemical potentials and masses (free meson gas), while in the square bracket are given the values of $\chi^2_{\rm dof}$ obtained with effective meson chemical potentials only.
Moreover, in the last line of the Tables is reported the best fit at the beam energy $\sqrt{s_{NN}}=130$ GeV corresponding to the same nine particle ratios and experimental data considered in Refs. \cite{braun01,zschie} ($N_{\rm dof}=7$),  and analyzed in Ref. \cite{chiapparini} within an analogue relativistic mean field model but with a constant (free parameter) effective meson mass ratio ($m_i^*/m_i=1.3$).

In Tables \ref{table_summary_tm1} and \ref{table_summary_gm3}, for both EOSs, we have reported the results for two different values of $x_{\sigma\Delta}$. As discussed in the previous Section, the lower values ($x_{\sigma\Delta}=1.25$ for TM1 and $x_{\sigma\Delta}=1.40$ for GM3) correspond to the case of not metastable $\Delta$-matter and, the upper values ($x_{\sigma\Delta}=1.33$ for TM1 and $x_{\sigma\Delta}=1.50$ for GM3) to the maximum couplings in which the metastable state can be realized in each EOS. In order to make a comparison between the two EOSs, we choose therefore two different couplings which correspond to a comparable situation for what concern the formation of $\Delta$-matter.

\begin{table*}
\caption{\label{table_summary_tm1} Summary of the results of the fits using
the particle ratios extracted from Refs. \cite{andronic06,star_prc09} (see text for details) for the TM1
parameter set. For comparison, in the round brackets are reported the respective values obtained without effective meson chemical potentials and masses (free gas of mesons). In the square brackets are given the values of $\chi^2_{\rm dof}$ obtained with effective meson chemical potentials only (no effective masses). In the last
line, the asterisk means that the fits are related to the same nine
particle ratios and experimental data considered in Refs. \cite{braun01,zschie,chiapparini}. }
\vspace{0.5cm}
\begin{center}
\begin{tabular}{c|ccc|ccc}
\hline\hline
 $\sqrt{s_{NN}}$
 &
 &$x_{\sigma\Delta}=1.25$
 &
 &
 &$x_{\sigma\Delta}=1.33$
 &
  {\vspace{0. cm}} \\
\cline{2-7}
(GeV) & $T$ (MeV) & $\mu_B$ (MeV) & $\chi^2_{\rm dof}$ &
$T$ (MeV) & $\mu_B$ (MeV) & $\chi^2_{\rm dof}$
 {\vspace{0.cm}} \\
\hline
2.70 & 51  (49) & 743 (741) & 1.19 (1.09) [1.11]& 51 (49) & 744 (741) & 1.24 (1.09) [1.12]\\
3.32 & 83  (68) & 656 (666) & 1.38 (2.43) [2.73]& 84 (68) & 651 (666) & 1.53  (2.44) [2.74]\\
3.84 & 101 (78) & 596 (620) & 1.63 (2.96) [3.04]& 94 (78) & 597 (619) & 2.78 (2.99) [3.06]\\
4.30 & 105 (86) & 563 (589) & 3.11 (2.76) [2.88]& 100 (86) & 564 (588) & 5.19  (2.80) [2.79]\\
4.85 & 107 (106) & 534 (551) & 1.90 (11.34) [11.24]& 107 (105) & 526 (549) & 2.80 (11.04) [11.11]\\
8.76 & 125 (121) & 405 (500) & 2.66 (7.87) [8.02]& 125 (120) & 393 (495) & 3.64  (7.26) [7.83]\\
12.3 & 133 (128) & 346 (456) & 2.98 (16.64) [16.82]& 132 (126) & 333 (460) & 2.84  (15.70) [15.88]\\
17.3 & 137 (131) & 306 (431) & 1.66 (13.61) [12.97]& 136 (130) & 291 (427) & 1.33  (12.76) [12.47]\\
62.4 & 153 (123) & 97 (49) & 1.67 (17.66) [17.71]& 152 (123) & 93 (49) & 0.39  (17.65) [17.66]\\
130 & 154 (128) & 46 (23) & 1.27 (15.61) [15.60]& 152 (128) & 42 (23) & 0.26  (15.58) [15.55]\\
200 & 154 (146) & 34 (24) & 0.40 (9.57) [9.52]& 153 (147) & 32 (25) & 0.05  (9.39) [9.35]\\
\hline
130$^*$ & 154 (149) [149]& 48 (33)& 0.56 (4.22) [4.21]& 153 (149) [148]& 46 (33) [32]& 0.52 (4.14) [4.12]\\
\hline\hline
\end{tabular}
\end{center}
\end{table*}

\begin{table*}
\caption{\label{table_summary_gm3} The same of Table \ref{table_summary_tm1} for the GM3 parameter set. }
\vspace{0.5cm}
\begin{center}
\begin{tabular}{c|ccc|ccc}
\hline\hline
 $\sqrt{s_{NN}}$
 &
 &$x_{\sigma\Delta}=1.40$
 &
 &
 &$x_{\sigma\Delta}=1.50$
 &
  {\vspace{0. cm}} \\
\cline{2-7}
(GeV) & $T$ (MeV) & $\mu_B$ (MeV) & $\chi^2_{\rm dof}$ &
$T$ (MeV) & $\mu_B$ (MeV) & $\chi^2_{\rm dof}$
 {\vspace{0.cm}} \\
\hline
2.70 & 50 (49)& 744 (741)& 1.25 (1.15) [1.25]& 51 (49)& 744 (741)& 1.28 (1.16) [1.26] \\
3.32 & 78 (68)& 655 (664)& 1.07 (2.37) [2.38]& 84 (68)& 647 (664)& 0.76  (2.38) [2.39]\\
3.84 & 111 (78)& 573 (617)& 0.70 (2.96) [3.05]& 99 (77)& 582 (618)& 1.20  (2.98) [3.07]\\
4.30 & 112 (86)& 542 (585)& 1.33 (2.81) [3.06]& 104 (86)& 550 (584)& 2.96 (2.84) [3.08]\\
4.85 & 112 (92)& 516 (555)& 1.11 (15.60) [16.05]& 111 (92)& 509 (555)& 1.94  (15.63) [16.09]\\
8.76 & 130 (126)& 387 (504)& 1.68 (12.48) [12.52]& 130 (125)& 375 (493)& 2.37 (11.51) [11.52]\\
12.3 & 138 (132)& 331 (477)& 3.11 (22.67) [21.57]& 137 (131)& 317 (465)& 2.55 (21.20) [19.80]\\
17.3 & 142 (136)& 293 (440)& 2.04 (18.61) [18.81]& 141 (135)& 278 (435)& 1.42  (17.35) [17.45]\\
62.4 & 161 (123)& 92 (48)& 3.27 (17.94) [17.95]& 159 (123)& 86 (48)& 1.17  (17.93) [17.95]\\
130 & 161 (128)& 43 (22)& 2.88 (16.01) [16.03]& 159 (128)& 40 (22)& 0.91  (15.99) [16.01]\\
200 & 161 (143)& 32 (19)& 1.01 (10.82) [10.84]& 160 (144)& 30 (19)& 0.29  (10.74) [10.76]\\
\hline
130$^*$ & 161 (148)& 45 (27)& 0.59 (4.80) [4.92]& 160 (148)& 43 (27)& 0.40 (4.76) [4.85] \\
\hline\hline
\end{tabular}
\end{center}
\end{table*}

As expected the obtained values of $\chi^2_{\rm dof}$ are very high without considering effective meson masses (except for the case $\sqrt{s_{NN}}=2.7$ GeV for which a very low fraction of mesons takes place). Comparable high values of $\chi^2_{\rm dof}$ are also obtained by considering effective meson chemical potentials only.
On the other hand, we can see that the introduction of the effective meson masses play a crucial role with results in very good agreement for both EOSs. It is interesting to observe that, for both EOSs, there is a better agreement with the lower values of the coupling ratio $x_{\sigma\Delta}$ at AGS energies, comparable values of $\chi^2_{\rm dof}$ with the lower and the higher values of $x_{\sigma\Delta}$ at SPS energies and a better agreement with the higher values of $x_{\sigma\Delta}$ at RHIC energies. This behavior is principally due to the fact that at high temperature and low baryon chemical potential the effective interaction is driven by the finite value of the $\sigma$-meson field, which significantly affects the value of the effective meson masses. This effect appears to be enhanced at higher values of $x_{\sigma\Delta}$ and reflects the fact that the $\Delta$-isobars dynamics results to be strongly influenced by the nuclear medium \cite{weise}. On the other hand, a higher value of $x_{\sigma\Delta}$ at higher beam energies could be interpreted as a phenomenological consequence of the increasing relevance of resonance matter in regime of high temperatures and low baryon chemical potentials \cite{zabrodin}.


From Tables \ref{table_summary_tm1} and \ref{table_summary_gm3}, the energy dependence of the two free parameters, $\mu_B$ and $T$, exhibits a monotonic dependence on the collision energy and appears very similar to the statistical thermal model results where $\mu_B$ decreases all the way up to RHIC energies, while $T$ rises rather sharply up to the lower SPS energies,  $\sqrt{s_{NN}}\approx 10$ GeV, and reaches afterwards almost constant values at RHIC energies.

Using the obtained values of $\mu_B$ and $T$ at different beam energies, by means the EOS, it is possible to obtain other relevant thermodynamic variables. In particular, it appears very interesting to find out the behavior of underlying dynamic quantities, such as the net baryon density $\rho_B$ and the energy density $\epsilon$, which are of more direct relevance to the collision dynamics respect to $\mu_B$ and $T$, because they are subject to corresponding conservation laws \cite{randrup2006}.

\begin{figure}[htb]
\begin{center}
\resizebox{0.5\textwidth}{!}{%
\includegraphics{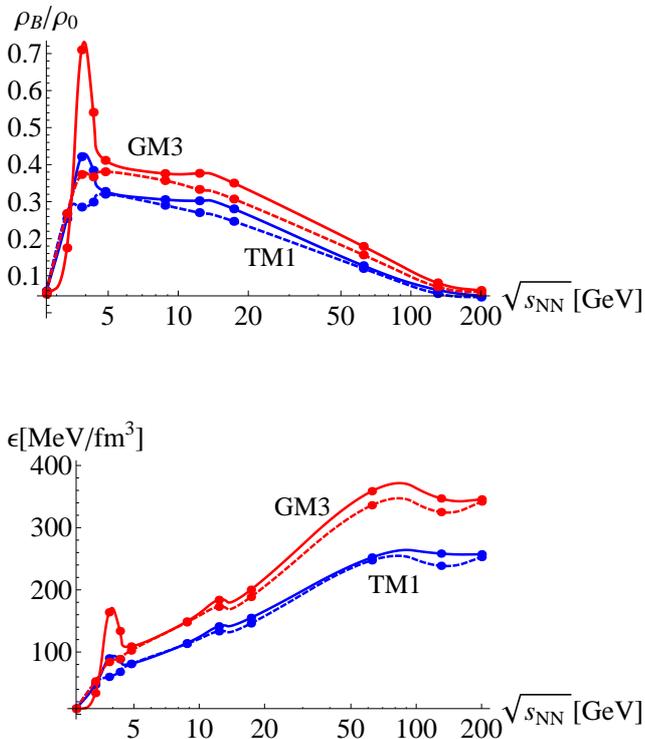}}
\caption{The energy dependence of the net baryon density $\rho_B$ (upper panel) and the energy density $\epsilon$ (lower panel) corresponding to the values of $\mu_B$
and $T$ reported in Tables \ref{table_summary_tm1} and \ref{table_summary_gm3}. The lower points correspond to the TM1 parameter set with $x_{\sigma\Delta}=1.25$ (solid lines) and $x_{\sigma\Delta}=1.33$ (dashed lines). The higher points correspond to the GM3 parameter set with $x_{\sigma\Delta}=1.40$ (solid lines) and $x_{\sigma\Delta}=1.50$ (dashed lines). The lines are strictly to help guide the eye.} \label{rhobeps}
\end{center}
\end{figure}

In order to avoid parametrization on the energy dependence of baryon chemical potential and temperature (since small uncertainties in $\mu_B$ and $T$ can lead to significant variation in $\rho_B$ and $\epsilon$ or smooth out possible sharp variations in the EOS), in Fig. \ref{rhobeps}, for the two EOSs, the net baryon density $\rho_B$ (upper panel) and the energy density $\epsilon$ (lower panel) are reported for different beam energies, corresponding to the values $\mu_B$ and $T$ of  Tables \ref{table_summary_tm1} and \ref{table_summary_gm3}.
The lines are strictly there to help guide the eye.

\begin{figure}[htb]
\begin{center}
\resizebox{0.5\textwidth}{!}{%
\includegraphics{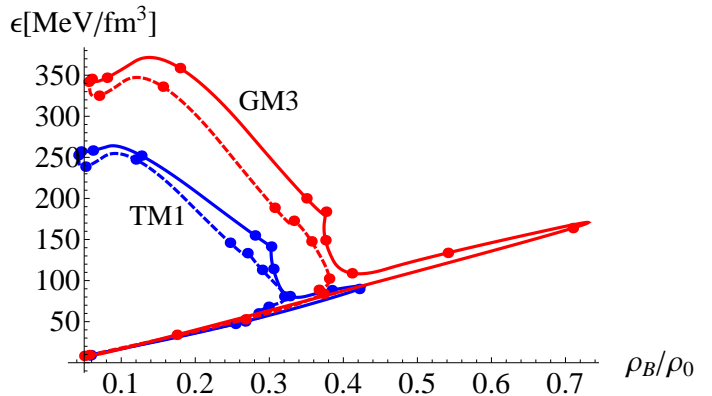}}
\caption{The same of Fig. \ref{rhobeps} for the energy density versus the net baryon density.}. \label{rhobepstris}
\end{center}
\end{figure}

First, we observe that, at fixed beam energy, with comparable values of the baryon chemical potential, greater values of baryon density and energy density are reached for the GM3 parameter set compared to the TM1 one. This aspect is principally due to a sensibly different value of the compression modulus $K$ for the two parameter sets.
In agreement with statistical thermal model predictions \cite{randrup2006}, for both EOSs, not only the net baryon density but also the energy density exhibits a non-monotonic dependence on the collision energy. The baryon density has an absolute maximum at about $\sqrt{s_{NN}}\approx 4$ GeV (which corresponds approximately to $\mu_B\approx 570\div 590$ MeV, depending on the parameter set) and at higher energies it decreases as a result of nuclear transparency. On the other hand, the energy density has a relative maximum at about the same beam energy. Both maximums tend to become less pronounced or to disappear by increasing the value of $x_{\sigma\Delta}$. Such non-monotonic trends appear to be emphasized in the $\rho_B-\epsilon$ plane, reported in Fig. \ref{rhobepstris}, with a sudden variation in the EOS, especially for lower value of $x_{\sigma\Delta}$. These abrupt changes in the EOS (which would become more smoothed with a parametrization on the energy dependence of $\mu_B$ and $T$) may be relevant in the planning of future compressed baryonic matter experiments \cite{senger,henning,bleicher}.

\vspace{0.5cm}

\section{Conclusion}

It is well known that mesons play a crucial role in the hadronic EOS especially in the regime of low baryon density and high temperature. Such particles are strongly interacting in the
environment with the other particles. In this paper, we have studied an effective relativistic mean field model with the inclusion of the full octet of baryons, the $\Delta$-isobars and the lightest pseudoscalar and vector mesons, by requiring the global conservation of baryon number, electric charge fraction and zero net strangeness.
The meson degrees of freedom have been incorporated in the EOS as a quasi-particle Bose gas with an effective meson chemical potential $\mu^*$ and an effective mass $m^*$ expressed in terms of
the $\sigma$, $\omega$, $\rho$ meson fields, responsible for the self-consistent mean field interaction.

The effective meson chemical potentials have been obtained from
a difference between the effective baryon chemical potentials respecting the
Gibbs conditions. In analogy, the effective meson masses have been expressed as a difference of the effective baryon masses respecting the strong interaction and on the basis of the main processes of the meson production/absorption involving different baryons. Such a phenomenological assumption implies a recipe in which the vacuum meson masses are reduced (enhanced) if the ratio $x_{\sigma B}=g_{\sigma B}/g_{\sigma N}$, between the $\sigma$-meson field coupling with the heavier baryon ($\Delta$ or hyperon particles involved in the meson production/absorption) and the nucleon one, is greater (lower) than one. A variation of the effective meson masses in-medium simulates in our simple scheme the relevance of meson-meson and meson-baryon self-interaction in the nuclear medium at finite temperatures and baryon densities.
Although the above assumptions have a very simple phenomenological nature, which cannot be extrapolated to any range of temperature and density, they have the noticeable advantage of not introducing additional parameters or couplings.

In this framework the hadron yield ratios have been studied and compared with the experimental ratios measured in central heavy ion collisions from AGS to RHIC energy range. Whenever possible, the same set of independent yield ratios adopted in Refs. \cite{star_prc09,qm2011} has been considered in the fitting procedure, in which temperature and baryon chemical potential are free parameters of the EOS. The results appear to be in very good overall accordance with the experimental ratios for the both used parameter sets (TM1 and GM3).
We have found that the presence of $\Delta$-isobar degrees of freedom play a crucial role in our scheme, in particular we have seen that a better agreement with the experimental data can be reached with lower values of the coupling ratio $x_{\sigma\Delta}$ (corresponding to a not metastable $\Delta$-isobar state) at lower beam energies and with higher values of $x_{\sigma\Delta}$ (corresponding to the formation of a $\Delta$-isobar metastable state) at higher beam energies.
From a phenomenological point of view, the coupling ratio $x_{\sigma\Delta}$ could be interpreted as a parameter which takes into account in an effective manner the formation of resonance states, very relevant in regime of high temperatures and low baryon chemical potentials.
Finally, we have seen that for both considered EOSs, in correspondence of the values of $T$ and $\mu_B$ extracted from the experimental data, the net baryon density and the energy density have a non-monotonic behavior with a maximum at a fixed value of the collision energy.

It is proper to remember that the values of temperature and baryon chemical potential, obtained from the fitting procedure with the experimental data, cannot properly be considered as the freeze-out parameters without taking into account particle decays, rescattering and annihilation effects. On the other hand, a correct reproduction of all particle production yields lies outside the scope of this paper. The main purpose of such a phenomenological comparison with experimental data is to provide useful indications about the relevance of the meson-baryon and meson-meson interaction in regime of finite values of temperature and baryon chemical potential where meson degrees of freedom become progressively more important.

\vspace{0.5cm}
\noindent
{\bf Acknowledgments}\\
It is a pleasure to thank F. Becattini, A. Drago, G. Garbarino and G. Pagliara for valuable suggestions and discussions and A. Andronic for details about the experimental data of Ref. \cite{andronic06}.

\end{document}